# Oxygen Content and Valence of Ru in $RuSr_2(Gd_{0.75}Ce_{0.25})_2Cu_2O_{10-d}$ (Ru-1222) Magnetosuperconductor

V.P.S. Awana[1], M. Karppinen[1], H. Yamauchi[1], M. Matvejeff[1], R.S. Liu[2] and L.-Y. Jang[3]

[1]Materials and Structures Laboratory, Tokyo Institute of Technology, Yokohama 226-8503, Japan

[2]Department of Chemistry, National Taiwan University, Taipei, Taiwan, R.O.C.

[3]Synchrotron Radiation Research Center, Hsinchu, Taiwan, R.O.C.

## Abstract

The valence of Ru was analyzed for two $RuSr_2(Gd_{0.75}Ce_{0.25})_2Cu_2O_{10-\delta}$ samples with different oxygen contents by Ru $L_{III}$–edge x-ray absorption near-edge structure (XANES) spectroscopy. For the sample as-synthesized in 1 atm $O_2$ the DC magnetization data measured in an applied field of 5 Oe showed a clear branching of the zero-field-cooled (ZFC) and field-cooled (FC) curves at around 100 K, a cusp at 85 K and a diamagnetic transition around 30 K in the ZFC part. A further confirmation for the superconductivity at 30 K was obtained from a resistance *vs.* temperature measurement. Annealing the as-synthesized sample in 100-atm $O_2$ atmosphere at 420 $^{o}$C increased the diamagnetic transition temperature from 30 K to 43 K. According to a thermogravimetric analysis, the oxygen content increased accordingly by *ca.* 0.15 oxygen atoms *per* formula unit. Quantitative analysis of the XANES spectra using $Sr_2RuO_4$ ($Ru^{IV}$) and $Sr_2GdRuO_6$ ($Ru^{V}$) as reference materials revealed a valence value of +4.74 and +4.81 for Ru in the as-synthesized and the 100-atm $O_2$-annealed sample, respectively. The obtained result suggests that the valence of Ru in Ru-1222 is affected by the change in oxygen content.

## 1. Introduction

According to a long-term common sense superconductivity and magnetic long-range order do not mutually exist within a single (thermodynamical) phase. The topic has been widely discussed in condensed matter physics over decades. Nevertheless, coexistence of high-$T_c$ superconductivity and magnetism was first reported for a rutheno cuprate of the Ru-1222 type, *i.e.* $RuSr_2(Gd_{0.75}Ce_{0.25})_2Cu_2O_{10-\delta}$ [1], and later for $RuSr_2GdCu_2O_{8-\delta}$ (Ru-1212) [2]. These reports have renewed the interest in the possible coexistence of superconductivity and magnetism. It is believed that in rutheno cuprates the $RuO_6$ octahedra in the charge reservoir are mainly responsible both for magnetism and for doping holes into the superconductive $CuO_2$ plane. This underlines the importance of the $RuO_6$ octahedra and the Ru valence state in these compounds. Despite the fact that various physical property measurements were carried out on Ru-1212 and Ru-1222 [3-8], no final consensus has been reached, *i.e.* discussion on their basic characteristics in terms of the carrier concentrations, doping mechanism, oxygen stoichiometry and the valence state of Ru has not been completed yet.

Structurally Ru-1222 with a layer sequence of SrO-$RuO_{2-\delta}$-SrO-$CuO_2$-(Gd,Ce)-$O_2$-(Gd,Ce)-$CuO_2$ is categorized as an $M_mA_{2k}B_sCu_{1+k}O_{m+4k+2s\pm\delta}$ or $M$-$m(2k)s(1+k)$ phase at $m$ = 1, $k$ = 1, $s$ = 2, *i.e.* it belongs to "category-B" of multi-layered copper oxides [9]. For $RuSr_2(Gd_{1-x}Ce_x)_2Cu_2O_{10-\delta}$ samples of the Ru-1222 phase loaded with oxygen under high oxygen pressure of 100 atm it has been shown by means of x-ray absorption near-edge structure (XANES) spectroscopy that Ru remains essentially pentavalent upon varying the Ce-substitution level within $0.3 \leq x \leq 0.5$ [8]. Here we report, however, based on Ru $L_{III}$-edge XANES spectroscopy that the valence of Ru is affected by changes in the oxygen content.

## 2. Experimental

Samples of composition $RuSr_2(Gd_{0.75}Ce_{0.25})Cu_2O_{10-\delta}$ were synthesized through a solid-state reaction route from $RuO_2$, $SrO_2$, $Gd_2O_3$, $CeO_2$ and CuO. Calcinations on the mixed powders were carried out at 1000, 1020, 1040 and 1060 $^oC$ each for 24 hours with intermediate grindings. The pressed bar-shaped pellets were annealed in a flow of oxygen at 1075 $^oC$ for 40 hours and subsequently cooled slowly over a span of another 20 hours down to room temperature. This sample is termed as "as-synthesized". Part of the as-synthesized sample was further annealed in high-pressure oxygen (100 atm) at 420 $^oC$ for 100 hours and subsequently cooled slowly to room temperature. This sample is termed as "100-atm $O_2$-annealed".

To confirm the phase purity and determine the lattice parameters x-ray diffraction patterns were recorded (MAC Science: MXP18VAHF[22]; Cu$K_\alpha$ radiation). Thermogravimetric (TG) analyses (Perkin Elmer: System 7) were carried out in a 5 % $H_2$/95 % atmosphere at the rate of 1 $^0C$/min to investigate the oxygen non-stoichiometry. Magnetization measurements were performed with a SQUID magnetometer (Quantum Design: MPMS-XL). Resistivity measurements were made in the temperature range of 5 to 300 K using a four-point-probe technique.

The Ru $L_{III}$-edge XANES measurements were performed at room temperature for polycrystalline samples at the BL15B beamline of the Taiwan Synchrotron Radiation Research Center (SRRC) in Hsinchu, using a Si(111) double-crystal monochomator. The energy resolution was ~0.47 eV at the Ru $L_{III}$ edge at 2838 eV. The spectra were recorded in a fluorescence mode using a Lytle detector. The $L_{II}$ and $L_{III}$ edges of Mo and Pd foils were used to calibrate the photon energy before and after the measurements. For the background subtraction the AUTOBK code [10] was employed in the measured energy range from ~200 eV below the edge to ~130 eV above the edge.

## 3. Results and Discussion

Both the as-synthesized and the 100-atm $O_2$-annealed Ru-1222 samples were of high quality in terms of phase purity. An x-ray diffraction pattern for the as-synthesized sample is shown in Fig. 1. The lattice parameters were determined from the diffraction data in the tetragonal space group *I4/mmm*: $a = b = 3.8337(6)$ Å and $c = 27.4926(9)$ Å for the as-synthesized sample, and $a = b = 3.8327(7)$ Å and $c = 27.3926(8)$ for the 100-atm $O_2$-annealed sample. The shorter lattice parameters for the 100-atm $O_2$-annealed sample manifested the fact that it was more completely oxygenated than the as-synthesized sample.

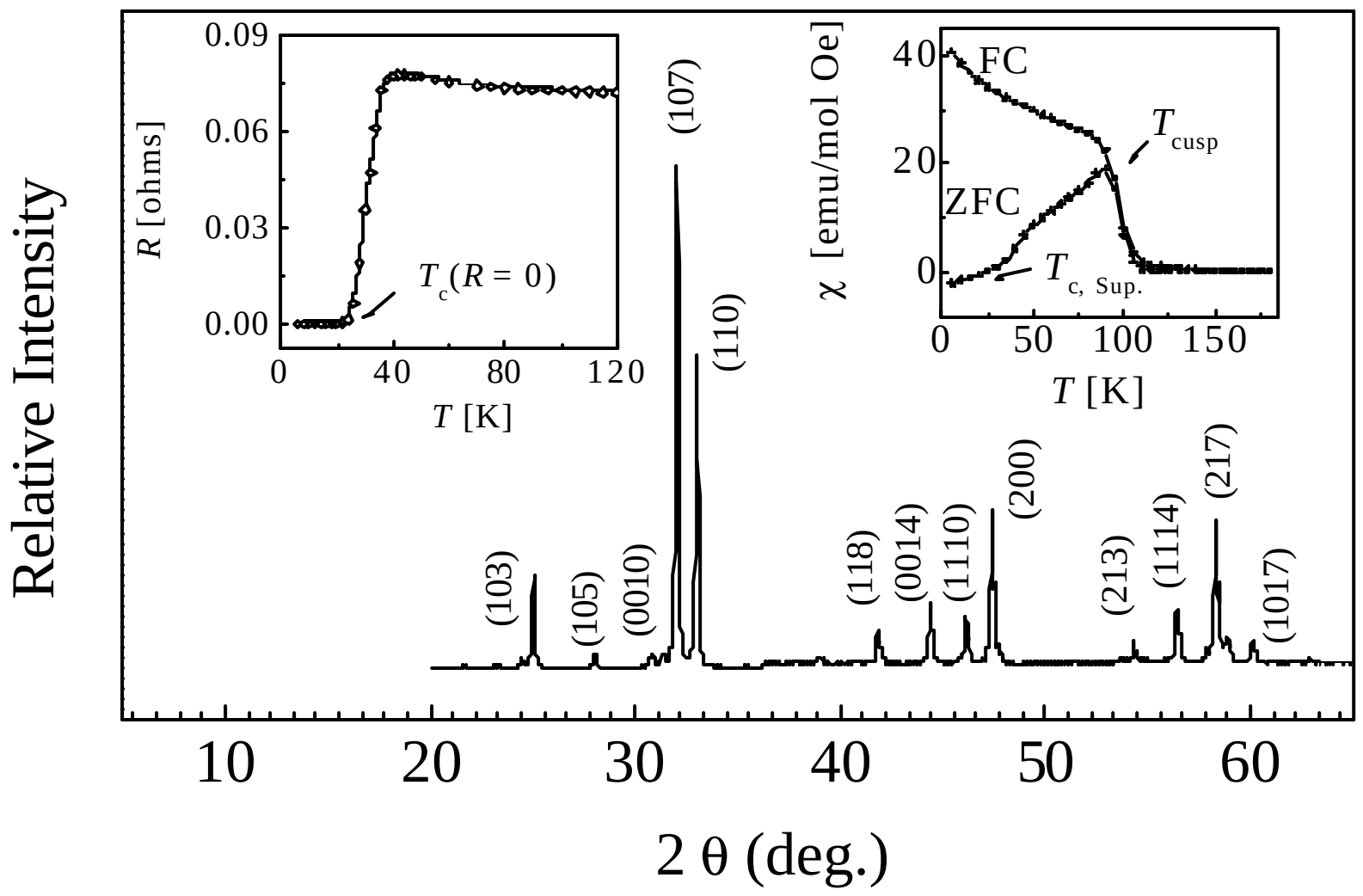

**Fig. 1.** X-ray diffraction pattern recorded for the as-synthesized Ru-1222 sample. The insets show the magnetic susceptibility (χ) *vs*. temperature (*T*) (right) and the resistance (*R*) *vs*. *T* (left) behaviours for the same sample.

The right inset of Fig. 1 shows the magnetic susceptibility *vs.* temperature behaviour in the temperature range of 5 to 160 K for the as-synthesized sample under an applied field of 5 Oe, measured in both zero-field-cooled (ZFC) and field-cooled (FC) modes. The ZFC and FC curves start branching around 140 K with a sharp upward turn for both around 100 K. The ZFC branch further shows a cusp at 85 K and a diamagnetic transition around 30 K. This is in agreement with earlier reports [1], and approves the "magneto-superconductive" nature of the presently studied sample. Superconductivity is also seen in electrical transport measurements on the same sample with $T_c$ ($R$ = 0) at 21 K (see the left inset in Fig.1). For the 100-atm $O_2$-annealed sample the $T_c$ ($R$ = 0) value increased to 43 K. This suggests that the as-synthesized sample was in an underdoped state.

From TG curves recorded upon heating in $H_2$/Ar it was seen that the complete reduction to a mixture of Cu metal, Ru metal and oxides of Sr, Gd and Ce occurred in parallel manner for the two samples. However, a difference in the total weight loss was observed for the two samples, reflecting a difference in their initial oxygen contents. From several repeated experiments, the difference in oxygen content of the as-synthesized ($T_c \approx 30$ ) and the 100-atm $O_2$-annealed ($T_c \approx$ 43 K) was established at ~0.15 oxygen atoms *per* formula unit.

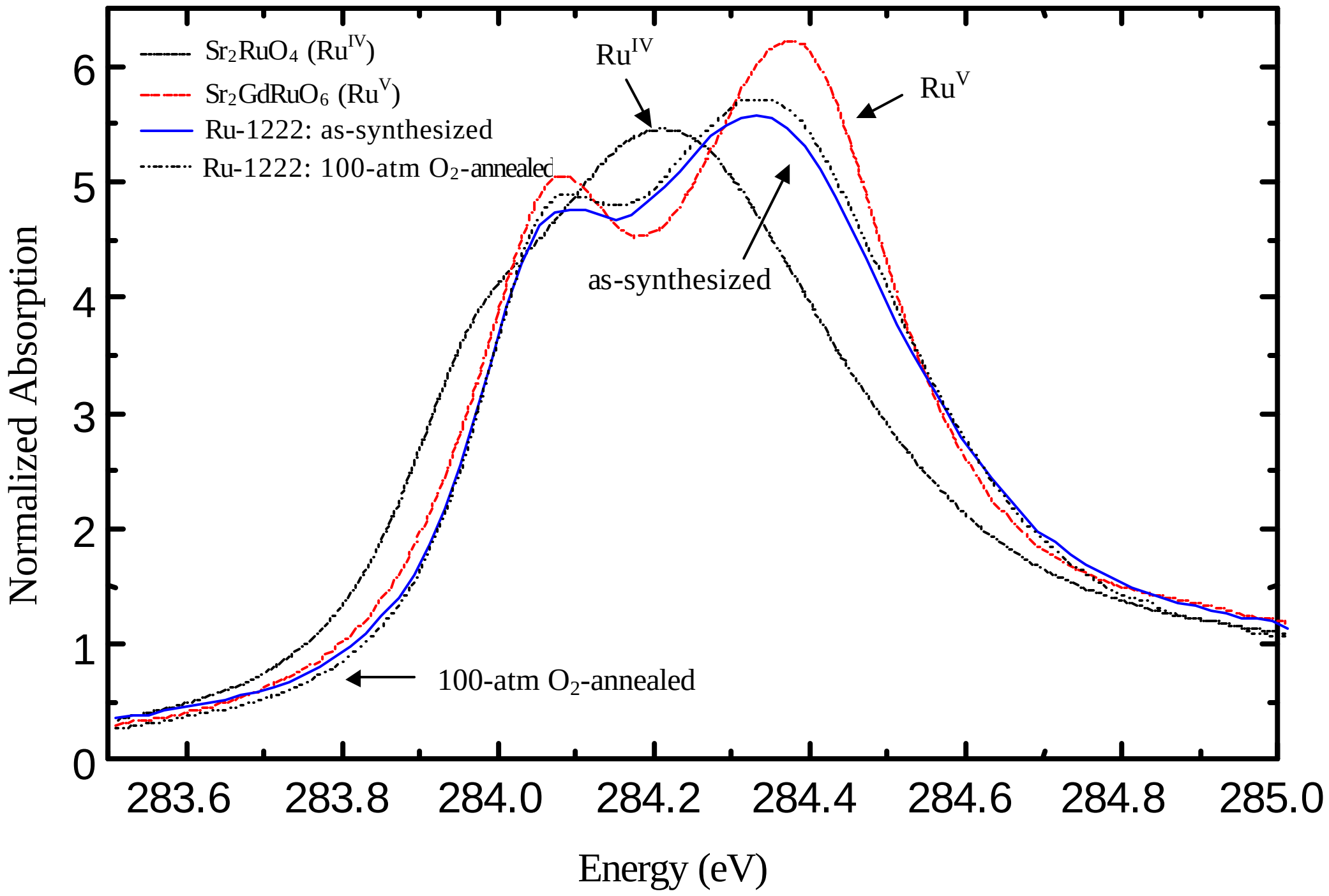

**Fig. 2.** Ru $L_{III}$-edge XANES spectra for the as-synthesized and the 100-atm $O_2$-annealed Ru-1222 samples. Also shown are the spectra for reference samples $Sr_2RuO_4$ ($Ru^{IV}$) and $Sr_2GdRuO_6$ ($Ru^{V}$).

The Ru $L_{III}$-edge XANES spectra for the two samples are displayed in Fig. 2. The spectra were analyzed quantitatively by fitting them to certain linear combinations of those for reference materials $Sr_2RuO_4$ ($Ru^{IV}$) and $Sr_2GdRuO_6$ ($Ru^{V}$). All the spectra exhibited two peaks, the higher-energy one and the lower-energy one being due to $2p \rightarrow e_g$ and $2p \rightarrow t_{2g}$ transitions, respectively [7,8,11]: with increasing Ru valence from +IV to +V, the crystal-field splitting increases and thereby the separation between the two peaks enhances. Furthermore, the peaks are accordingly shifted by ~1.5 eV to the higher energy. From Fig. 2, both the Ru-1222 samples are between the two reference materials in terms of the Ru valence. Fitting the spectra revealed a valence value of +4.74 for the as-synthesized sample and +4.81 for the 100-atm $O_2$-annealed sample.

The obtained result suggested that the valence of Ru in Ru-1222 was affected by the change in oxygen content. It was useful to compare the presently obtained Ru valence values to that reported for a $RuSr_2(Gd_{0.7}Ce_{0.3})_2Cu_2O_{10-\delta}$ sample (+4.95) with $T_c \approx 60$ K [8]. For the three samples the Ru valence ($T_c$) values thus were: +4.74 (30 K), +4.81 (43 K) and +4.95 (60 K). (Note that the XANES measurements and analyses were carried out in parallel ways for all the three samples.) The latter two samples were both annealed under 100 atm oxygen pressure, but with different temperature programs. It was thus likely that the one previously reported [8] had a somewhat higher oxygen content than the present 100-atm $O_2$-annealed sample.